\documentclass[twocolumn,twoside]{article}
\usepackage{graphics}

\newcommand{\bpartial}{\mbox{\boldmath $\partial$}}

\newcommand{\bgamma}{\mbox{\boldmath $\gamma$}}

\newcommand{\mgamma}{\mbox{$\gamma$}}

\begin{document}
\title{HOT DENSE MAGNETIZED SPINOR MATTER IN PARTICLE AND ASTROPARICLE PHYSICS: THE ROLE OF BOUNDARIES}
\author{Yu.A.\,Sitenko\\[2mm]
Bogolyubov Institute for Theoretical Physics,
National Academy of Sciences of Ukraine,\\
14-b Metrologichna Street, 03680 Kyiv, Ukraine
}
\date{}

\maketitle

ABSTRACT.
We study the influence of boundaries on chiral effects in hot dense relativistic spinor matter in a strong magnetic field which is 
orthogonal to the boundaries. The most general set of boundary conditions ensuring the confinement of matter within the boundaries 
is employed. We find that the chiral magnetic effect disappears, whereas the chiral separation effect stays on, becoming dependent on 
temperature and on a choice of boundary conditions. As temperature increases from zero to large values, a stepped-shape behaviour 
of the chiral separation effect as a function of chemical potential is changed to a smooth one. A choice of the boundary condition 
can facilitate either amplification or diminution of the chiral separation effect; in particular, the effect can persist even at 
zero chemical potential, if temperature is finite. This points at a significant role of boundaries for physical systems with hot 
dense magnetized spinor matter, i.e. compact astrophysical objects (neutron stars and magnetars), relativistic heavy-ion 
collisions, novel materials known as the Dirac and Weyl semimetals.\\[1mm] 
\\[2mm]

{\bf 1. Introduction}\\[1mm]

Properties of hot dense spinor matter in a strong magnetic field are intensively studied during the last decade. An interest 
to this subject is driven by diverse areas of contemporary physics, ranging from 
particle and astroparticle physics to cosmology and even condensed-matter physics. Relativistic heavy-ion collisions 
(Kharzeev D.E.: 2014), compact astrophysical objects (neutron stars and magnetars) (Ferrario L., Melatos A., and Zrake J.: 2015), 
the early universe (Tashiro H., Vachaspati T., and Vilenkin A.: 2012), novel materials known as the Dirac and Weyl semimetals 
(Vafek O. and Vishwanath A.: 2014) are the main physical systems where such studies are relevant. In the case of temperature, 
chemical potential, and the inverse magnetic length exceeding considerably the mass of a relativistic quantized spinor matter field, 
it has been shown in theory (Metlitski M.A. and Zhitnitsky A.R.: 2005; Fukushima K., Kharzeev D.E., and Warringa H.J.: 2008) that 
persistent and nondissipative currents emerge in thermal equilibrium, resulting in a variety of chiral effects in hot dense 
magnetized matter; see review in Miransky V.A. and Shovkovy I.A.: 2015, and references therein.   
 
So far chiral effects were mostly considered in unbounded (infinite) matter, which may be relevant for cosmological applications, 
perhaps. For all other applications (to particle, astroparticle, and condensed matter physics), an account has to be taken of the 
finiteness of physical systems, and the role of boundaries in chiral effects in bounded matter has to be clearly exposed. The 
concept of quantized matter fields which are confined to bounded spatial regions is quite familiar in the context of condensed 
matter physics: collective excitations (e.g., spin waves and phonons) exist only inside material samples and do not spread outside. 
Nevertheless, a quest for boundary conditions ensuring the confinement of quantized matter was initiated in particle physics, in 
the context of a model description of hadrons as bags containing quarks (Bogolioubov P.N.: 1968; Chodos A., Jaffe R.L., Johnson K., 
Thorn C. B., and Weisskopf V.: 1974). Motivations for a concrete form of the boundary condition may differ in detail, but the key 
point is that the boundary condition has to forbid any flow of quark matter across the boundary, see Johnson K.: 1975. However, 
from this point of view, the bag boundary conditions proposed in Bogolioubov P.N.: 1968, and in Chodos A., Jaffe R.L., Johnson K., 
Thorn C. B., and Weisskopf V.: 1974 are not the most general ones. It has been rather recently realized that the most general 
boundary condition ensuring the confinenent of relativistic quantized spinor matter within a simply connected boundary involves 
four arbitrary parameters (Akhmerov A.R. and Beenakker C.W.J.: 2008;  Al-Hashimi M.H. and Wiese U.-J.: 2012), and the explicit form 
of such a condition has been given (Sitenko Yu.A.: 2015; Sitenko Yu.A. and Yushchenko S.A.: 2015; Sitenko Yu.A.: 2016a). To study an 
impact of the background magnetic field on confined matter, one has to choose the magnetic field configuration with respect to the 
boundary surface. The primary interest is to understand the effect of a boundary which is transverse to the magnetic field strength 
lines. Then the simplest geometry is that of a slab in the uniform magnetic field directed perpendicular. It should 
be noted that such a geometry can be realized in condensed matter physics by putting slices of Dirac or Weyl semimetals in an 
external transverse magnetic field. Note also that the slab geometry is conventional in a setup for the Casimir effect 
(Casimir H.B.G.: 1948); see review in Bordag M., G. L. Klimchitskaya G. L., Mohideen U., and Mostepanenko V. M.: 2009. 

As a first step toward the full theory of chiral effects in bounded matter, the chiral effects in dense magnetized 
ultrarelativistic spinor matter at zero temperature in a slab were considered in publication Gorbar E.V., Miransky V.A., 
Shovkovy I.A., and Sukhachov P.O.: 2015, with the use of the bag boundary condition of Bogolioubov P.N.: 1968. Our aim is to extend 
the consideration to the case of nonzero temperature and the most general boundary condition.
\\[2mm]

{\bf 2. Thermal equilibrium for chiral spinor matter}\\[1mm]

We start with the operator of the second-quantized spinor field in a static background,
\begin{equation}
\hat{\Psi}({\bf r},t)=\sum\limits_{E_\lambda>0}e^{-{\rm i}E_\lambda t}\left\langle {\bf r}|\lambda\right\rangle\hat{a}_\lambda
+\sum\limits_{E_\lambda<0}e^{-{\rm i}E_\lambda t}\left\langle {\bf r}|\lambda\right\rangle\hat{b}_\lambda^{\dagger},\label{eq1}
\end{equation}
where $\hat{a}_\lambda^{\dagger}$ and $\hat{a}_\lambda$ ($\hat{b}_\lambda^{\dagger}$ and $\hat{b}_\lambda$) are the spinor particle (antiparticle) creation and destruction operators satisfying  anticommutation relations,
\begin{equation}
\left[\hat{a}_\lambda,\,\hat{a}_{\lambda'}^{\dagger}\right]_+ = \left[\hat{b}_\lambda,\,\hat{b}_{\lambda'}^{\dagger}\right]_+
=\left\langle \lambda|\lambda'\right\rangle, \label{eq2}
\end{equation}
and $\left\langle {\bf r}|\lambda\right\rangle$ is the solution to the stationary Dirac equation,
\begin{equation}
H\left\langle {\bf r}|\lambda\right\rangle=E_\lambda\left\langle {\bf r}|\lambda\right\rangle,\label{eq3}
\end{equation}
$H$ is the Dirac Hamiltonian, $\lambda$ is the set of parameters (quantum numbers) specifying a one-particle state, 
$E_\lambda$ is the energy of the state; wave functions $\left\langle {\bf r}|\lambda\right\rangle$ satisfy the requirement of 
orthonormality 
\begin{equation}
\int\limits_{\Omega}{\rm d}^3r\left\langle \lambda|{\bf r}\right\rangle\left\langle {\bf r}|\lambda'\right\rangle
=\left\langle \lambda|\lambda'\right\rangle\label{eq4}
\end{equation}
and completeness
\begin{equation}
\sum\left\langle {\bf r}|\lambda\right\rangle\left\langle \lambda|{\bf r}'\right\rangle=I\delta({\bf r}-{\bf r}');\label{eq5}
\end{equation}
summation is over the whole set of states, and $\Omega$ is the quantization volume. 

Conventionally, the operators of dynamical variables (physical observables) in second-quantized theory are defined as bilinears 
of the fermion field operator (1). One can define the fermion number operator,
\begin{eqnarray}
\hat{N}=\frac{1}{2}\int\limits_{\Omega}{\rm d}^3r(\hat{\Psi}^{\dagger}\hat{\Psi}-\hat{\Psi}^T\hat{\Psi}^{\dagger T})\nonumber \\ 
= \sum\left[\hat{a}_\lambda^{\dagger}\hat{a}_\lambda-\hat{b}_\lambda^{\dagger}\hat{b}_\lambda
-\frac{1}{2}{\rm sgn}(E_\lambda)\right], \label{eq6}
\end{eqnarray}
and the energy (temporal component of the energy-momentum vector) operator,
\begin{eqnarray}
\hat{P}^0=\frac{1}{2}\int\limits_{\Omega}{\rm d}^3r(\hat{\Psi}^{\dagger}H\hat{\Psi}-\hat{\Psi}^TH^T\hat{\Psi}^{\dagger T})
\nonumber \\ 
= \sum|E_\lambda|\left(\hat{a}_\lambda^{\dagger}\hat{a}_\lambda+\hat{b}_\lambda^{\dagger}\hat{b}_\lambda
-\frac{1}{2}\right),\label{eq7}
\end{eqnarray}
where superscript $T$ denotes a transposition and ${\rm sgn}(u)$ is the sign function [${\rm sgn}(\pm u)=\pm 1$ at $u>0$]. 
The average of operator $\hat{U}$ over the grand canonical ensemble is defined as (see, e.g., Das A.: 1997)
\begin{equation}
\left\langle \hat{U}\right\rangle_{T,\mu}=\frac{{\rm Sp} \, \hat{U}{\rm exp}\left[-(\hat{P}^0
-\mu\hat{N})/T\right]}{{\rm Sp} \, {\rm exp}\left[-(\hat{P}^0-\mu\hat{N})/T\right]}, \label{eq8}
\end{equation}
where equilibrium temperature $T$ is defined in the units of the Boltzmann constant, chemical potential is denoted by $\mu$, 
and ${\rm Sp}$ denotes the trace or the sum over the expectation values in the Fock state basis created by operators in (2).
Let us take operator $\hat{U}$ in the form
\begin{equation}
\hat{U}=\frac{1}{2}\left(\hat{\Psi}^{\dagger}\Upsilon\hat{\Psi}-\hat{\Psi}^T\Upsilon^T\hat{\Psi}^{\dagger T}\right),\label{eqno9}
\end{equation}
where $\Upsilon$ is an element of the Dirac-Clifford algebra. The explicit form of $\hat{U}$, $\hat{P}^0$ and $\hat{N}$ in terms 
of the creation and destruction operators is inserted in (9); then one obtains
\begin{equation}
\left\langle \hat{U}\right\rangle_{T,\mu}=
-\frac{1}{2}{\rm tr}\left\langle {\bf r}|\Upsilon\tanh [(H-\mu I)(2T)^{-1}]|{\bf r}\right\rangle,\label{eq10}
\end{equation}
where ${\rm tr}$ denotes the trace over spinor indices. We are considering the quantized charged spinor field in the background of 
a static uniform magnetic field with strength ${\bf B}={\bpartial}\times{\bf A}$, where ${\bf A}$ is the vector 
potential of the magnetic field. Assuming that the magnetic field is strong (supercritical) and ultrarelativistic spinor matter 
is at high temperature and high density, 
\begin{equation}
|eB|>>m^2, \quad T>>m, \quad |\mu|>>m, \label{eq11}
\end{equation}
we shall neglect the mass of the spinor matter field, putting $m=0$ in the following. Thus the Dirac Hamiltonian takes form 
\begin{equation}
H=-{\rm i}\gamma^0{\bgamma}\cdot({\bpartial}-{\rm i}e{\bf A}), \label{eq12}
\end{equation}
where $e$ is the charge of the matter field and natural units $\hbar=c=1$ are used. Owing to the presence of chiral symmetry, 
\begin{equation}
 [H, \gamma^5]_- = 0, \label{eq13}
\end{equation}
where $\gamma^5=-{\rm i}\gamma^0\gamma^1\gamma^2\gamma^3$ ($\gamma^0$, $\gamma^1$, $\gamma^2$, and $\gamma^3$ are the generating 
elements of the Dirac-Clifford algebra, and $\gamma^5$ is defined according to Okun L.B.: 1982), one can introduce also the 
following average:
\begin{equation}
\left\langle \hat{U}\right\rangle_{T,{\mu}_5}=\frac{{\rm Sp} \, \hat{U}{\rm exp}\left[-(\hat{P}^0
-{\mu}_5\hat{N}^{5})/T\right]}{{\rm Sp} \, {\rm exp}\left[-(\hat{P}^0-{\mu}_5\hat{N}^{5})/T\right]}, \label{eq14}
\end{equation}
where 
\begin{equation}
\hat{N}^{5}=\frac{1}{2}\int\limits_{\Omega}{\rm d}^3r(\hat{\Psi}^{\dagger}\gamma^5\hat{\Psi}-
\hat{\Psi}^T\gamma^{5T}\hat{\Psi}^{\dagger T}) \label{eq15}
\end{equation}
is the axial charge and ${\mu}_5$ is the chiral chemical potential. For operator $\hat{U}$ in the form of (9), one obtains
\begin{equation}
\left\langle \hat{U}\right\rangle_{T,{\mu}_5}=-\frac{1}{2}{\rm tr}\left\langle {\bf r}|\Upsilon\tanh [(H-
{\mu}_5 \gamma^5)(2T)^{-1}]|{\bf r}\right\rangle, \label{eq16}
\end{equation}

Thus, there are two types of thermal averaging, when chiral symmetry is present. For instance, one can define the vector current 
density as either an average over the standard grand canonical ensemble, 
\begin{equation}
{\bf J}=\left\langle \hat{U}\right\rangle_{T,\mu}\biggr|_{\Upsilon={\mgamma}^0{\bgamma}}, \label{eq17}
\end{equation}
or an average over the grand canonical ensemble with the chiral chemical potential, 
\begin{equation}
{\bf J}=\left\langle \hat{U}\right\rangle_{T,{\mu}_5}\biggr|_{\Upsilon={\mgamma}^0{\bgamma}}. \label{eq18}
\end{equation}
As we shall see, there is no contradiction between these two different definitions of the same physical quantity, since at least 
one of them results in zero.  
\\[2mm]

{\bf 3. Chiral effects in the unnbounded space}\\[1mm]

A solution to the Dirac equation in the background of a static uniform magnetic field is well described in the literature, 
see, e.g., Akhiezer A.I. and Berestetskij V.B.: 1965. The one-particle energy spectrum in the case of the massless spinor field is
\begin{eqnarray}
E_{nk}=\pm\omega_{nk}, \quad \omega_{nk}=\sqrt{2n|eB|+k^{2}},\nonumber \\
-\infty<k<\infty, \quad n=0,1,2,...\, ,\label{19}
\end{eqnarray}
$k$ is the value of the wave number vector along the magnetic field, and $n$ numerates the Landau levels. A straightforward 
calculation of (18) with the use of the explicit form of spinor wave functions of definite chiralities immediately yields 
\begin{equation}
{\bf J}\equiv\left\langle \hat{U}\right\rangle_{T,{\mu}_5}\biggr|_{\Upsilon={\mgamma}^0{\bgamma}}=
-\frac{e{\bf B}}{2\pi^2}{\mu}_5,   \label{eq20}
\end{equation}
whereas (17) turns out to be zero. Similarly, one can calculate the axial charge density, 
\begin{eqnarray}
J^{05}\equiv\left\langle \hat{U}\right\rangle_{T,\mu,{\mu}_5}\biggr|_{\Upsilon={\mgamma}^5} = 
\frac{|eB|}{2\pi^2}\Biggl\{ {\mu}_5 \Biggr.
\nonumber \\
+\sum\limits_{n=1}^{\infty}\int\limits_{0}^{\infty}{\rm d}k\,\frac{{\rm sinh}[({\mu}_5 + \mu)/T]}{{\rm cosh}[({\mu}_5 + \mu)/T]
+{\rm cosh}(\sqrt{2n|eB| + k^2}/T)}\nonumber \\
\Biggl. +\sum\limits_{n=1}^{\infty}\int\limits_{0}^{\infty}{\rm d}k\,
\frac{{\rm sinh}[({\mu}_5 - \mu)/T]}{{\rm cosh}[({\mu}_5 - \mu)/T]
+{\rm cosh}(\sqrt{2n|eB| + k^2}/T)}\Biggr\},  \label{eq21}
\end{eqnarray}
and the axial current density,
\begin{equation}
{\bf J}^5\equiv\left\langle \hat{U}\right\rangle_{T,\mu}\biggr|_{\Upsilon={\mgamma}^0{\bgamma}{\mgamma}^5} = 
-\frac{e{\bf B}}{2\pi^2}\mu. \label{eq22}
\end{equation}
Note that ${\bf J}$ (20) and ${\bf J}^5$ (22) are known as the chiral magnetic (Fukushima K., Kharzeev D.E., and Warringa H.J.: 
2008) and chiral separation (Metlitski M.A. and Zhitnitsky A.R.: 2005) effects, respectively. Only the lowest Landau level ($n=0$) 
contributes to (20) and (22), and both effects are of topological origin, being related to a quantum anomaly, see 
Alekseev A.Y., Cheianov V.V., and Frohlich J.: 1998; Giovannini M. and Shaposhnikov M.: 1998; Son D.T. and 
Zhitnitsky A.R.: 2004; Kharzeev D.E., McLerran L.D., and Warringa H.J.: 2008; Basar G. and Dunne G.V.: 2013. 
Defining the left and right current densities, 
\begin{equation}
{\bf J}_L = \frac12 \left({\bf J}+{\bf J}^5\right)  \label{eq23}
\end{equation}
and
\begin{equation}
{\bf J}_R = \frac12 \left({\bf J}-{\bf J}^5\right),  \label{eq24}
\end{equation}
as well as the left and right chemical potentials, 
\begin{equation}
{\mu}_L = \mu + {\mu}_5 \label{eq25}
\end{equation}
and
\begin{equation}
{\mu}_R = \mu - {\mu}_5, \label{eq26}
\end{equation}
one can rewright (20) and (22) as
\begin{equation}
{\bf J}_L =-\frac{e{\bf B}}{4\pi^2}{\mu}_L  \label{eq27}
\end{equation}
and
\begin{equation}
{\bf J}_R =\frac{e{\bf B}}{4\pi^2}{\mu}_R.   \label{eq28}
\end{equation}
Actually, the chiral magnetic effect was first discovered in the form of (27) in 
Vilenkin A.: 1980 (with a missing extra factor of 1/2).
\\[2mm]

{\bf 4. Chiral effects in a slab}\\[1mm]

To study an influence of a background magnetic field on the properties of hot dense spinor matter, one has to account for the 
fact that the realistic physical systems are bounded. Our interest is in an effect of the static magnetic field with strength lines 
which are orthogonal to a boundary. Then, as was already noted, the simplest geometry of a material sample is that of a straight 
slab in the uniform magnetic field directed perpendicular. The one-particle energy spectrum in this case is [cf. (19)]
\begin{equation}
E_{nl}=\pm\omega_{nl},\,\,\,\,\omega_{nl}=\sqrt{2n|eB|+k_l^2},\,\,\,\,n=0,1,2,\ldots,\label{eq29}
\end{equation} 
where the values of the wave number vector along the magnetic field, $k_l$, are to be determined by the 
boundary condition.

The most general boundary condition ensuring the confinement of relativistic spinor matter within a simply connected boundary 
is (see Sitenko Yu.A. and Yushchenko S.A.: 2015; Sitenko Yu.A.: 2016a)
\begin{eqnarray}
\left\{I-\gamma^0\left[{\rm e}^{{\rm i}\varphi\gamma^5}\cos\theta  \right. \right. \nonumber \\
\left. \left. + (\gamma^1\cos\varsigma + 
\gamma^2\sin\varsigma)\sin\theta\right]{\rm e}^{{\rm i}\tilde{\varphi}{\mgamma}^0({\bgamma}\cdot\mathbf{n})}\right\} \nonumber \\
\times \chi(\mathbf{r})\left.\right|_{\mathbf{r}\in \partial\Omega}=0, \label{eq30}
\end{eqnarray}
where $\mathbf{n}$ is the unit normal to surface $\partial\Omega$ bounding spatial region $\Omega$ 
and $\chi(\mathbf{r})$ is the confined spinor matter wave function, $\mathbf{r} \in \Omega$; 
matrices $\gamma^1$ and $\gamma^2$ in (30) are chosen to obey condition
\begin{equation}
[\gamma^1,\,{\bgamma}\cdot\mathbf{n}]_+=
[\gamma^2,\,{\bgamma}\cdot\mathbf{n}]_+=[\gamma^1,\,\gamma^2]_+=0,\label{eq31}
\end{equation}
and the boundary parameters in (30) are chosen to vary as
\begin{eqnarray}
-\frac{\pi}{2}<\varphi\leq\frac{\pi}{2}, \quad -\frac{\pi}{2}\leq\tilde{\varphi}<\frac{\pi}{2}, \nonumber \\
0\leq\theta<\pi, \quad 0\leq\varsigma<2\pi. \label{eq32}
\end{eqnarray}
The MIT bag boundary condition (Johnson K.: 1975),
\begin{equation}
(I+{\rm i}{\bgamma}\cdot\mathbf{n})\chi(\mathbf{r})\left.\right|_{\mathbf{r}\in \partial\Omega}=0, \label{eq33}
\end{equation}
is obtained from (30) at $\varphi=\theta=0$, $\tilde{\varphi}=-{\pi}/{2}$.

The boundary parameters in (30) can be interpreted as the self-adjoint extension parameters. The self-adjointness of the 
one-particle energy (Dirac Hamiltonian in the case of relativistic spinor matter) operator in first-quantized theory is required 
by general principles of comprehensibility and mathematical consistency; see, e.g., Bonneau G., Faraut J., and Valent G.: 2001. 
To put it simply, a multiple action is well defined for a self-adjoint operator only, allowing for the construction of functions 
of the operator, such as resolvent, evolution, heat kernel and zeta-function operators, with further implications upon second 
quantization.

In the case of a disconnected boundary consisting of two simply connected components, 
$\partial\Omega=\partial\Omega^{(+)}\bigcup\partial\Omega^{(-)}$, there are in general eight boundary parameters: 
$\varphi_{+}$, $\tilde{\varphi}_{+}$, $\theta_+$, and $\varsigma_+$ corresponding to $\partial\Omega^{(+)}$; and 
$\varphi_{-}$, $\tilde{\varphi}_{-}$, $\theta_{-}$, and $\varsigma_-$ corresponding to $\partial\Omega^{(-)}$. If spatial region $\Omega$ has the 
form of a slab bounded by parallel planes, $\partial\Omega^{(+)}$ and $\partial\Omega^{(-)}$, separated by distance $a$, then the boundary condition takes form  
\begin{equation}
\left(I-K^{(\pm)}\right)
\chi(\mathbf{r})\left.\right|_{z=\pm a/2}=0,\label{eq34}
\end{equation}
where 
\begin{eqnarray}
K^{(\pm)}=\gamma^0\left[{\rm e}^{{\rm i}\varphi_{\pm}\gamma^5}\cos\theta_{\pm} \right. \nonumber \\
\left.+ (\gamma^1\cos\varsigma_{\pm} + 
\gamma^2\sin\varsigma_{\pm})\sin\theta_{\pm}\right]{\rm e}^{\pm{\rm i}\tilde{\varphi}_{\pm}\gamma^0\gamma^z}, \label{eq35}
\end{eqnarray}
coordinates $\mathbf{r}=(x,\,y,\,z)$ are chosen in such a way that $x$ and $y$ are tangential to the boundary, while $z$ is 
normal to it, and the position of $\partial\Omega^{(\pm)}$ is identified with $z=\pm a/2$. The confinement of matter inside the 
slab means that the vector bilinear, $\chi^{\dag}(\mathbf{r})\gamma^0{\gamma}^z\chi(\mathbf{r})$, vanishes at the slab boundaries,
\begin{equation}
\chi^{\dag}(\mathbf{r})\gamma^0{\gamma}^z\chi(\mathbf{r})\left.\right|_{z=\pm a/2}=0,\label{eq36}
\end{equation}
and this is ensured by condition (34). As to the axial bilinear, 
$\chi^{\dag}(\mathbf{r})\gamma^0{\gamma}^z\gamma^5\chi(\mathbf{r})$, it vanishes at the slab boundaries,
\begin{equation}
\chi^{\dag}(\mathbf{r})\gamma^0{\gamma}^z\gamma^5\chi(\mathbf{r})\left.\right|_{z=\pm a/2}=0,\label{eq37}
\end{equation}
in the case of $\theta_+ = \theta_- = \pi/2$ only, that is due to relation 
\begin{equation}
[K^{(\pm)}\left.\right|_{\theta_{\pm} = \pi/2}, \gamma^5]_- = 0. \label{eq38}
\end{equation}
However, note that, as a massless spinor particle is reflected from an impenetrable boundary, its helicity is flipped. 
Since the chirality equals plus or minus the helicity, chiral symmetry has to be necessarily broken by the confining boundary 
condition. Thus the case of $\theta_+ = \theta_- = \pi/2$ is not acceptable on the physical grounds. Moreover, there is a symmetry 
with respect to rotations around a normal to the slab, and the cases differing by values of 
$\varsigma_+$ and $\varsigma_-$ are physically indistinguishable, since they are related by such a rotation. The only way to 
avoid the unphysical degeneracy of boundary conditions with different values of $\varsigma_+$ and $\varsigma_-$ is to fix 
$\theta_+=\theta_-=0$. Then $\chi^{\dag}(\mathbf{r})\gamma^0{\gamma}^z\gamma^5\chi(\mathbf{r})$ is nonvanishing at the slab 
boundaries, and the boundary condition takes form
\begin{equation}
\left\{I-\gamma^0\exp\left[{\rm i}\left(\varphi_\pm\gamma^5\pm\tilde{\varphi}_\pm\gamma^0\gamma^z\right)\right]\right\}
\chi(\mathbf{r})\left.\right|_{z=\pm a/2}=0. \label{eq39}
\end{equation}
Condition (39) determines the spectrum of the wave number vector in the $z$ direction, $k_l$. The requirement that this spectrum 
be real and unambiguous yields constraint (see Sitenko Yu.A. and Yushchenko S.A.: 2015; Sitenko Yu.A.: 2016a)
\begin{equation}
\varphi_+=\varphi_-=\varphi, \quad \tilde{\varphi}_+=\tilde{\varphi}_-=\tilde{\varphi};\label{eq40}
\end{equation}
then the $k_l$ spectrum is determined implicitly from relation 
\begin{equation}
k_l\sin\tilde{\varphi}\cos(k_l a)+(E_{... l}\cos\tilde{\varphi}-m\cos\varphi)\sin(k_l a)=0, \label{eq41}
\end{equation}
where $m$ is the mass of the spinor matter field and $E_{... l}$ is the energy of the one-particle state.
In the case of the massless spinor matter field, $m=0$, and the background uniform magnetic field perpendicular to the slab,
${\bf B}=(0,0,B)$, $E_{... l}$ takes the form of $E_{n l}$ (29), and relation (41) is reduced to
\begin{equation}
k_l\sin\tilde{\varphi}\cos(k_l a)+E_{n l}\cos\tilde{\varphi}\sin(k_l a)=0, \label{eq42}
\end{equation}
depending on one parameter only, although the boundary condition depends on two parameters,
\begin{equation}
\left\{I-\gamma^0\exp\left[{\rm i}\left(\varphi\gamma^5 \pm  \tilde{\varphi}\gamma^0\gamma^z\right)\right]\right\}
\chi(\mathbf{r})|_{z=\pm a/2}=0. \label{eq43}
\end{equation}

Using the explicit form of standing waves with the $k_l$ spectrum determined by (42), one can compute the 
averages of type (8). As to the averages of type (14), they cannot be defined, because the confining boundary condition 
necessarily  breaks chiral symmetry: standing waves inside a slab are formed from counterpropagating waves of opposite chiralities.
It is straightforward to check the validity of relations
\begin{equation}
{\bf J}=J^{05}=0. \label{eq44}
\end{equation}
Similarly, the components of the axial current density, which are orthogonal to the direction of the magnetic field, vanish as well,
\begin{equation}
J^{x5}=J^{y5}=0.\label{eq45}
\end{equation}
As to the component of the axial current density, which is along the magnetic field, only the lowest Landau level ($n=0$) 
contributes to it, and hence the $z$ component of the axial current density is
\begin{eqnarray}
J^{z5}=-\frac{eB}{2\pi a}\left[\sum\limits_{k_l^{(+)}>0}f_+(k_l^{(+)}) - \sum\limits_{k_l^{(-)}>0}f_-(k_l^{(-)}) \right. \nonumber \\ 
\left. - \frac{1}{2}\sum\limits_{k_{l}^{(+)}>0}1 + \frac{1}{2}\sum\limits_{k_{l}^{(-)}>0}1\right], \label{eq46}
\end{eqnarray}
where
\begin{equation}
f_\pm(k)=\left[e^{(k\mp\mu)/T}+1\right]^{-1}, \label{eq47}
\end{equation}
and the two last sums which are independent of temperature and chemical potential correspond to the contribution of the vacuum 
fluctuations. The calculation of the sums over $l$ yields (see Sitenko Yu.A.: 2016b, 2016c)
\begin{eqnarray}
J^{z5}=-\frac{eB}{2\pi a} \nonumber \\
\times \Biggl\{{\rm sgn}(\mu)F\Biggl(|\mu|a + {\rm sgn}(\mu)\left[\tilde{\varphi}-
{\rm sgn}(\tilde{\varphi}){\pi}/{2}\right];Ta\Biggr) \Biggr. \nonumber \\
\Biggl. - \frac{1}{\pi}\left[\tilde{\varphi}-
{\rm sgn}(\tilde{\varphi}){\pi}/{2}\right]\Biggr\} , \label{eq48}
\end{eqnarray}
where 
\begin{eqnarray}
F(s;t)=\frac{s}{\pi} \nonumber \\
-\frac{1}{\pi}\int\limits_{0}^{\infty}{{\rm d}v\,\frac{\sin(2s){\rm sinh}(\pi/t)}
{[\cos(2s)+{\rm cosh}(2v)][{\rm cosh}(\pi/t)+\cos(v/t)]}}\nonumber \\ 
+\frac{{\rm sinh} \left\{[{\rm arctan} ({\rm tan} s)]/t\right\}}{{\rm cosh}[\pi/(2t)]+{\rm cosh}
\left\{[{\rm arctan} ({\rm tan} s)]/t\right\}}.\label{eq49}
\end{eqnarray}

In view of relation
\begin{equation}
\lim\limits_{a\rightarrow \infty}\frac{1}{a}F(|\mu|a;Ta)=|\mu|/{\pi},\label{eq50}
\end{equation}
the case of a magnetic field filling the whole (infinite) space (Metlitski M.A. and Zhitnitsky A.R.: 2005) is obtained from (48) 
as a limiting case [cf. (22)],
\begin{equation}
\lim\limits_{a\rightarrow \infty}J^{z5}=-\frac{eB}{2\pi^2} \mu.\label{eq51}
\end{equation}
Unlike this unrealistic case, the realistic case of a magnetic field confined to a slab of finite width is temperature dependent, 
see (48) and (49). In particular, we get
\begin{eqnarray}
\lim\limits_{T\rightarrow 0}J^{z5}=-\frac{eB}{2\pi a} \nonumber \\ 
\times \Biggl[{\rm sgn}(\mu)\left[\!\!\left[\frac{|\mu|a+
{\rm sgn}(\mu)\tilde{\varphi}}{\pi} 
+ \Theta(-\mu\tilde{\varphi})\right]\!\!\right] \Biggr. \nonumber \\ 
\Biggl.- \frac{\tilde{\varphi}}{\pi} + 
\frac{1}{2}{\rm sgn}(\tilde{\varphi})\Biggr] \label{eq52}
\end{eqnarray}
and
\begin{equation}
\lim\limits_{T\rightarrow \infty}J^{z5}=-\frac{eB}{2\pi^2} \mu; \label{eq53}
\end{equation} 
here $[\![u]\!]$ denotes the integer part of quantity $u$ (i.e. the integer which is less than or equal 
to $u$), and $\Theta(u)=\frac{1}{2}[1+{\rm sgn}(u)]$ is the step function. As follows from (48), the boundary condition that is 
parametrized by $\tilde{\varphi}$ can serve as a source which is additional to the spinor matter density: the contribution of the 
boundary to the axial current effectively enhances or diminishes the contribution of the chemical potential. Because of the boundary 
condition, the chiral separation effect can be nonvanishing even at zero chemical potential,
\begin{eqnarray}
J^{z5}|_{\mu=0}=-\frac{eB}{2\pi a}\Biggl\{F(\tilde{\varphi}-{\rm sgn}(\tilde{\varphi})\pi/2; Ta) \Biggr. \nonumber \\ 
\Biggl. - \frac{1}{\pi}\left[\tilde{\varphi}-{\rm sgn}(\tilde{\varphi}){\pi}/{2}\right]\Biggr\}; \label{eq54}
\end{eqnarray}
the latter vanishes in the limit of infinite temperature, 
\begin{equation}
\lim\limits_{T \rightarrow \infty}J^{z5}|_{\mu=0}=0.\label{eq55}
\end{equation}
The trivial boundary condition, $\tilde{\varphi}=-\pi/2$, yields spectrum $k_l=(l+\frac 12)\frac{\pi}{a}$ 
$\quad$ ($l=0,1,2,\ldots$), which is the same in the setups of both bag models (Bogolioubov P.N.: 1968) and (Chodos A., 
Jaffe R.L., Johnson K., Thorn C. B., and Weisskopf V.: 1974). The axial current density at zero temperature for this case was 
obtained in Gorbar E.V., Miransky V.A., Shovkovy I.A., and Sukhachov P.O.: 2015,
\begin{equation}
J^{z5}\left.\right|_{T=0, \,\, \tilde{\varphi}=-\pi/2} 
= - \frac{eB}{2\pi a}{\rm sgn}(\mu)\left[\!\!\left[ \frac{|\mu|a}{\pi}+\frac{1}{2} \right]\!\!\right]. \label{eq56}
\end{equation}
The "bosonic-type" spectrum, $k_l=l\frac{\pi}{a}$ $\quad$ ($l=0,1,2,\ldots$), is yielded by 
$\tilde{\varphi}=0$, and, due to the contribution of the vacuum fluctuations, the axial current density is continuous at 
this point,
\begin{equation}
\lim\limits_{\tilde{\varphi}\rightarrow 0_+}J^{z5} = \lim\limits_{\tilde{\varphi}\rightarrow 0_-}J^{z5}. \label{eq57}
\end{equation}
\\[2mm]

{\bf 5. Summary and discussion}\\[1mm]

We have considered the influence of boundaries on chiral effects in hot dense magnetized relativistic spinor matter. An issue of the 
confining boundary condition plays the key role in this study. In the absence of boundaries there exist the chiral magnetic effect 
which is exhibited by the nondissipative vector current along the magnetic field, see (20), and the chiral separation effect which 
is exhibited by the nondissipative axial current in the same direction, see (22); both currents are temperature independent. As 
boundaries are introduced and the matter volume is shrinked to a slab which is transverse to the magnetic field, the fate of these 
currents is different. The axial current stays on, becoming dependent on temperature and on a choice of boundary conditions, see 
(48) and (49); as temperature increases from zero to large values, a stepped-shape behaviour of the axial current density as a function 
of chemical potential is changed to a smooth one, see (52) and (53). The vector current, as well as the axial charge, is extinct, 
see (44), because of boundary condition (36). Thus, the chiral magnetic effect in a slab is eliminated by the confining boundary 
condition.

Magnetic fields of the order of the QCD energy scale squared can be produced in the quark-gluon plasma created in relativistic 
heavy-ion collisions (as a result of electric currents from the colliding charged ions); see Kharzeev D.E.: 2014.
Magnetic fields up to $10^{15}$ Gauss may exist in some compact stars (magnetars); see Glendenning N.K.: 2000; Turola R, Zane S., 
and Watts A.L.: 2015. Even stronger fields are generated in progenitor magnetars during the core collapse after the supernovae 
explosion (Ferrario L., Melatos A., and Zrake J.: 2015). It has been claimed that the chiral magnetic effect (20) plays a 
significant role in these processes; see Boyarsky A., Ruchayskiy O., and Shaposhnikov M.: 2012; Sigl G. and Leite N.: 2016. Another 
claim is that the chiral magnetic effect is responsible for a solution of the problem of large kicks of pulsars (Charbonneau J. and 
Zhitnitsky A.: 2010; Charbonneau J., Hoffman, and Heyl J.: 2010). The latter problem is that pulsars exhibit rapid proper motion 
with velocities ranging from 100 to 1600 km/s, and about 15$\%$ of pulsars has velocities above 1000 km/s (Arzoumanian Z., 
Chernoff D.F., and Cordes J.M.: 2002). Large velocities are unambiguously confirmed with the model independent measurement of 
pulsar B1505+55 moving at $1083^{+103}_{-90}$ km/s (Chatterjee S., Vlemmings W.H.T., Brisken W.F., et al: 2005). Persistent 
topological current (20) has been proposed in the capacity of an engine which pushes a fast pulsar like a rocket (Charbonneau J. and 
Zhitnitsky A.: 2010).

However, if finiteness of a physical system in the direction of the magnetic field is taken into account, then the topological 
vector current, as well as the axial charge, disappears, that is due to the boundary condition confining spinor (electron, 
quark, or nucleon) matter inside the magnetar. This circumstance changes essentially the whole picture of the large kick mechanism.  
In view of the above, the only necessary ingredients are a strong magnetic field and the high density of the relativistic charged 
spinor matter in the core of a compact star (magnetar or protomagnetar). Such an environment develops axial currents with the 
specific temperature and chemical potential dependence which is almost stepped-shape in the case of degenerate matter 
($T \ll |\mu|$) and a somewhat smooth otherwise, see (48), (49), (52), and (53). The pertinent carriers of these currents are 
electrons in the case of the nuclear matter, or electrons and quarks in the case of the deconfined quark matter (in supposed quark 
stars). Since the currents are persistent, they are capable of delivering the asymmetry of neutrinos created by weak interaction 
from the interior of the star to its surface without dissipation, thus producing a proper motion of the star as a whole. The 
details of the mechanism are yet to be elaborated.\\[5mm]

{\it Acknowledgements.}
I would like to thank the Organizers of the XVI Odessa Gamow International Conference-School ``Astronomy and beyond: 
Astrophysics, Cosmology, Cosmomicrophysics, Astroparticle Physics, Radioastronomy and Astrobiology'' for kind 
hospitality during this interesting and inspiring meeting. The work was supported by the National Academy of Sciences
of Ukraine (Project No.0112U000054) and by the ICTP -- SEENET-MTP Grant PRJ-09 ``Strings and Cosmology''.
\\[3mm]
{\bf References\\[2mm]}
Akhiezer A.I. and Berestetskij V.B.: 1965, {\it Quantum Electrodynamics} (Interscience, New York).\\
Akhmerov A.R. and Beenakker C.W.J.: 2008, {\it  Phys. Rev. B} {\bf 77}, 085423.\\
Alekseev A.Y., Cheianov V.V., and Frohlich J.: 1998, {\it Phys. Rev. Lett.} {\bf 81}, 3503.\\
Al-Hashimi M.H. and Wiese U.-J.: 2012, {{\it Ann. Phys. (Amsterdam)} {\bf 327}, 1.\\ 
Arzoumanian Z., Chernoff D.F., and Cordes J.M.: 2002, {\it Astrophys. J.} {\bf 568}, 289.\\  
Basar G. and Dunne G.V.: 2013, {{\it Lect. Notes Phys.} {\bf 871}, 261.\\ 
Bogolioubov P.N.: 1968, {\it Ann. Inst. Henri Poincare A} \textbf{8}, 163.\\
Bonneau G., Faraut J., and Valent G.: 2001, {\it  Amer. J. Phys.} {\bf 69}, 322.\\
Bordag M., G. L. Klimchitskaya G. L., Mohideen U., and Mostepanenko V. M.: 2009,
{\it Advances in the Casimir Effect} (Oxford University Press, Oxford).\\
Boyarsky A., Ruchayskiy O., and Shaposhnikov M.: 2012, {\it Phys. Rev. Lett.} {\bf 109}, 111602.\\
Casimir H.B.G.: 1948, {\it  Proc. Kon. Ned. Akad. Wetenschap B} {\bf 51}, 793.\\
Charbonneau J., Hoffman, and Heyl J.: 2010, {\it  Mon. Not. Roy. Astron. Soc.} {\bf 404}, L119.\\
Charbonneau J. and Zhitnitsky A.: 2010, {\it  J. Cosmol. Astropart. Phys.} 08 (2010) 010.\\
Chatterjee S., Vlemmings W.H.T., Brisken W.F., et al: 2005, {\it Astrophys. J.} {\bf 630}, L61.\\  
Chodos A., Jaffe R.L., Johnson K., Thorn C. B., and Weisskopf V.: 1974, {\it  Phys. Rev. D} {\bf 9}, 3471.\\
\linebreak\vfill\pagebreak\noindent
Das A.: 1997, {\it Finite Temperature Field Theory} (World Scientific, Singapore).\\
Ferrario L., Melatos A., and Zrake J.: 2015, {\it Space Sci. Rev.} {\bf 191}, 77.\\
Fukushima K., Kharzeev D.E., and Warringa H.J.: 2008, {\it Phys. Rev. D} {\bf 78}, 074033.\\
Giovannini M. and Shaposhnikov M.: 1998, {\it Phys. Rev. D} \textbf{57}, 2186.\\
Glendenning N.K.: 2000, {\it Compact Stars: Nuclear Physics, Particle Physics, and General Relativity} (Spriger, New York).\\
Gorbar E.V., Miransky V.A., Shovkovy I.A., and Sukhachov P.O.: 2015, {\it Phys. Rev. B} \textbf{92}, 245440.\\
Johnson K.: 1975, {\it Acta Phys. Pol. B} {\bf 6}, 865.\\
Kharzeev D.E.: 2014, {\it Prog. Part. Nucl. Phys.} {\bf 75}, 133.\\
Kharzeev D.E., McLerran L.D., and Warringa H.J.: 2008, {\it Nucl. Phys. A} {\bf 803}, 227.\\
Metlitski M.A. and Zhitnitsky A.R.: 2005, {\it Phys. Rev. D} \textbf{72}, 045011.\\
Miransky V.A. and Shovkovy I.A.: 2015, {\it  Phys. Rep.} \textbf{576}, 1.\\
Okun L.B.: 1982, {\it Leptons and Quarks} (Elsevier Science Publishers B.V., Amsterdam).\\
Sigl G. and Leite N.: 2016, {\it  J. Cosmol. Astropart. Phys.} 01 (2016) 025.\\
Sitenko Yu.A.: 2015, {\it Phys. Rev. D} {\bf 91}, 085012.\\
Sitenko Yu.A.: 2016a, {\it J. Phys. Conf. Ser.} \textbf{670}, 012048.\\
Sitenko Yu.A.: 2016b, {\it Europhys. Lett.} {\bf 114}, 61001.\\
Sitenko Yu.A.: 2016c, {\it Phys. Rev. D} {\bf 94}, 085014.\\
Sitenko Yu.A. and Yushchenko S.A.: 2015, {\it Int. J. Mod. Phys. A} \textbf{30}, 1550184.\\
Son D.T. and Zhitnitsky A.R.: 2004, {\it  Phys. Rev. D} \textbf{70}, 074018.\\
Tashiro H., Vachaspati T., and Vilenkin A.: 2012, {\it  Phys. Rev. D} \textbf{86}, 105033.\\
Turola R, Zane S., and Watts A.L.: 2015, {\it Rep. Prog. Phys.} \textbf{78}, 116901.\\
Vafek O. and Vishwanath A.: 2014, {\it Annu. Rev. Condens. Matter Phys.} {\bf 5}, 83.\\
Vilenkin A.: 1980, {\it Phys. Rev. D} \textbf{22}, 3080.\\

\end{document}